\RequirePackage{amsmath}
\documentclass[10pt]{iopart}
\usepackage{iopams}
\usepackage[dvips]{graphicx}
\usepackage{cite}

\begin{document}

\title[]{A linear magnetic flux-to-voltage transfer function of differential DC SQUID}

\author{I I Soloviev$^{1,2,3,4}$, V I Ruzhickiy$^{1,3,4,5}$, N V Klenov$^{1,2,3,4,5}$, S V Bakurskiy$^{1,2,3,4}$ and M Yu Kupriyanov$^{1,2}$}

\address{$^{1}$Lomonosov Moscow State University Skobeltsyn Institute of Nuclear Physics, 119991, Moscow, Russia}
\address{$^{2}$Moscow Institute of Physics and Technology, State University, 141700 Dolgoprudniy, Moscow region, Russia}
\address{$^{3}$MIREA - Russian Technological University, 119454, Moscow, Russia}
\address{$^{4}$N. L. Dukhov All-Russia Research Institute of Automatics, 127055, Moscow, Russia}
\address{$^{5}$Physics Department, Moscow State University, 119991, Moscow, Russia}

\ead{isol@phys.msu.ru} \vspace{10pt}

\begin{abstract}
A superconducting quantum interference device with differential
output or ``DSQUID'' was proposed earlier for operation in the
presence of large common-mode signals. The DSQUID is the
differential connection of two identical SQUIDs. Here we show that
besides suppression of electromagnetic interference this device
provides effective linearization of DC SQUID voltage response. In
the frame of the resistive shunted junction model with zero
capacitance, we demonstrate that Spur-Free Dynamic Range (SFDR) of
DSQUID magnetic flux-to-voltage transfer function is higher than
SFDR~$>$~100~dB while Total Harmonic Distortion (THD) of a signal is
less than THD~$<$~$10^{-3}\%$ with a peak-to-peak amplitude of a
signal being a quarter of half flux quantum, $2\Phi_a = \Phi_0/8$.
Analysis of DSQUID voltage response stability to a variation of the
circuit parameters shows that DSQUID implementation allows doing
highly linear magnetic flux-to-voltage transformation at the cost of
a high identity of Josephson junctions and high-precision current
supply.
\end{abstract}


\vspace{2pc} \noindent{\it Keywords}: DC SQUID, voltage response,
linearity, working margins, DSQUID

%\submitto{\SUST}
\maketitle
\ioptwocol

\section{Introduction}

Modern Josephson junction fabrication technology \cite{Tolp,HYPNP}
allows the development of complex circuits \cite{OurBeil} with
high-precision control of their parameters. Both low-temperature and
high-temperature superconductor (LTS and HTS) technology provide a
possibility to fabricate SQUID arrays with the number of Josephson
junctions about a million \cite{TolpLarge,HTSSQIF}. This expands the
area of SQUID applications to the one where SQUID-based structures
should ideally act as a linear magnetic flux-to-voltage transformers
\cite{SQLin2,HTSSQIF,HTSSQIFA,HTSAnt,ADRHYP,IPHTSQ1,Brag}: from
electrically small antennas to analog-to-digital convertor circuits
and from susceptometers to SQUID-based multiplexers.

SQUID-based structures with high dynamic range and highly linear
voltage response obtained without a feedback loop were named the
``superconducting quantum arrays'' (SQA) \cite{SQA1,SQA2,SQA3}. Two
types of cells were proposed as basic blocks of SQA. These are the
bi-SQUID \cite{BSQ1,BSQ2,BSQ3,BSQ4,BSQ5,BSQ6,BSQ7,BSQ8} and the
so-called differential quantum cell (DQC)
\cite{SQLin2,SQA1,SQA2,SQA3,DQC1,DQC2,DQC3,DQC4,DQC5}.
Unfortunately, despite several attempts to realization of
bi-SQUID-based structures \cite{BSQ3,BSQex1,BSQex2} no outstanding
results were reported \cite{BSQ8}. DQC seems to deliver better
performance for SQA \cite{SQLin2}. However, since DQC is a
differential connection of identical parallel SQUID arrays, it
usually occupies a large area which is not convenient in some cases.

In this paper, we consider the simplest version of DQC - two
identical DC SQUIDs with a differential output which we call a
``DSQUID'', see Fig.~\ref{Fig1}a. Earlier it was shown that the
DSQUID allows obtaining high common-mode rejection ratio
\cite{diffSQUID}. This feature is especially useful where
SQUID-based system contains long wiring. It was also noted that the
effects of background magnetic fields and of temperature
fluctuations are also suppressed due to this differential
configuration. \cite{diffSQUID}.

Here we show that besides the presented advantages the DSQUID
possesses high voltage response linearity of DQC. We report the
range of DSQUID parameters providing high linearity of its voltage
response as well as analysis of the linearity decrease with
deviation of the circuit parameters from their optimal values.

\section{Model}

DSQUID voltage response is obtained by subtraction of voltage
responses of its parts: $u_\Sigma = u_+ - u_-$, see Fig.~\ref{Fig1}.
Josephson junctions of DSQUID ought to be overdamped to accomplish
the high linearity of DQC \cite{SQLin2}. Equality of DSQUID parts
naturally suggests the using of LTS technology where the
technological spread of parameters can be minimized. For the
temperature $T = 4.2$~K the effective current noise value is $I_T =
(2\pi/\Phi_0)k_B T \approx 0.18~\mu$A, where $\Phi_0$ is the
magnetic flux quantum and $k_B$ is the Boltzmann constant. The
choice of Josephson junction critical current $I_c \geq 180~\mu$A
leads to dimensionless noise intensity $\gamma = I_T/I_c \leq
10^{-3}$ which makes the noise impact to DQC characteristics to be
insignificant \cite{SQLin2}. Transfer function of each SQUID of
DSQUID is calculated in the frame of the well-known Resistive
Shunted Junction (RSJ) model with zero capacitance, accordingly. The
system of equations describing SQUID in terms of Josephson phase sum
and difference $\varphi_\pm = (\varphi_1 \pm \varphi_2)/2$ (where
$\varphi_{1,2}$ are Josephson phases of SQUID junctions) is as
follows:

\begin{figure}[]
\begin{center}
\includegraphics[bb = 0 0 503 293,width=1\columnwidth,keepaspectratio]{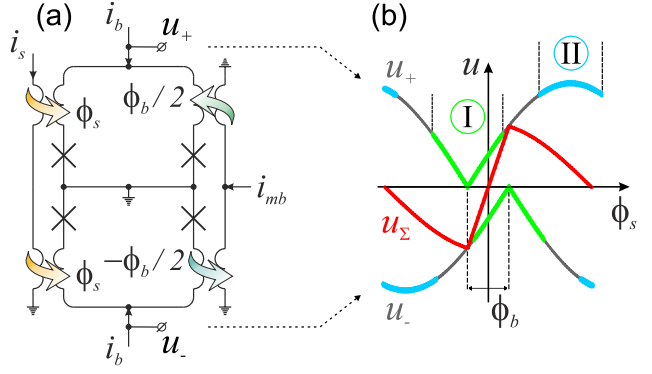}
\end{center}
\caption{(a) DSQUID scheme. $i_b$ is the bias current, $i_s$ is the
signal current providing the signal flux, $\phi_s$, and $i_{mb}$ is
the magnetic bias current providing the bias flux, $\phi_b$. DSQUID
voltage response $u_\Sigma$ is obtained between nodes $u_\pm$. (b)
Sketch of DSQUID voltage response formation as subtraction of
voltage responses of its parts: $u_\Sigma = u_+ - u_-$. DC SQUID
voltage response regions linearized in DSQUID are marked by numbers
I and II.} \label{Fig1}
\end{figure}

\begin{subequations} \label{SQUIDeq}
\begin{equation}
\frac{d\varphi_+}{d\tau} = i_b - \sin\varphi_+\cos\varphi_-,
\end{equation}
\begin{equation}
\frac{d\varphi_-}{d\tau} = -(\varphi_- - \phi_e)/\beta_l -
\sin\varphi_-\cos\varphi_+,
\end{equation}
\end{subequations}
where time $\tau = t\omega_c$ is normalized to the characteristic
frequency $\omega_c = 2\pi I_c R_n / \Phi_0$, $R_n$ is the junction
shunt resistance, $i_b = I_b/2I_c$ is the normalized bias current,
$\beta_l = \pi L I_c/\Phi_0$ is the normalized SQUID inductance, and
$\phi_e = \pi \Phi_e / \Phi_0$ is the normalized applied magnetic
flux. For each SQUID of DSQUID $\phi_e$ is the sum of the signal
flux, $\phi_s$, and the magnetic bias flux setting the working point
and the width of the working region, $\pm\phi_b/2$, as it is shown
in Fig.~\ref{Fig1}.

We use two approaches for estimation of voltage response linearity.
The first one is calculation of so-called Spur-Free Dynamic Range
(SFDR). In this approach we apply the external magnetic flux to each
SQUID of DSQUID in the form,
\begin{equation}
\phi_e = (\phi_a/2)\sin\omega_1 \tau + (\phi_a/2)\sin\omega_2 \tau
\pm\phi_b/2,
\end{equation}
where frequencies $\omega_2 = 1.1 \omega_1$ are much smaller than
Josephson oscillation frequency, $\omega_{1,2} \ll \omega_J$,
$\phi_a$ is the amplitude of signal. SFDR is calculated as a ratio
of one of the signal tones to maximum amplitude of distortions
arising in spectrum of output signal due to nonlinearity of magnetic
flux-to-voltage transfer function.

The second approach is calculation of Total Harmonic Distortion
(THD). In this case the applied signal contains only one harmonic
component:
\begin{equation}
\phi_e = \phi_a\sin\omega\tau \pm\phi_b/2.
\end{equation}
THD is calculated as THD~$= \sqrt{\sum_{n=2}^{\infty}A_n^2}/A_1$,
where $A_n$ are amplitudes of output signal spectral harmonics.

SFDR and THD calculation requires finding the accurate shape of
DSQUID voltage response by numerical solutions of system
(\ref{SQUIDeq}) for each SQUID. For this purpose we define the
dependence $\varphi_-(\varphi_+)$ combining equations
(\ref{SQUIDeq}a,b),
\begin{equation}
\frac{d\varphi_-}{d\varphi_+} = \frac{-(\varphi_- - \phi_e)/\beta_l
- \sin\varphi_-\cos\varphi_+}{i_b - \sin\varphi_+\cos\varphi_-},
\end{equation}
and then calculate the period of Josephson oscillations using
(\ref{SQUIDeq}a):
\begin{equation}
T = \int_0^{2\pi} \frac{d\varphi_+}{i_b -
\sin\varphi_+\cos\varphi_-(\varphi_+)}.
\end{equation}
This gives us the Josephson oscillation frequency, $\omega_J =
2\pi/T$, which is equal to time-averaged voltage, $u$, normalized to
$I_c R_n$ product.

Calculation of voltage response shape for estimation of magnetic
flux-to-voltage transfer coefficient is done much faster using
analytical expressions presented in \cite{SQLin1,SQAn}.

\section{Linearization}

Differential connection of two identical SQUIDs in DSQUID with their
mutual flux bias allows subtraction of some part of SQUID voltage
response from its mirrored image, see Fig.~\ref{Fig1}b, due to
symmetry and periodicity of SQUID voltage response. This subtraction
leads to partial compensation of nonlinear terms in DSQUID magnetic
flux-to-voltage transfer function for two regions of SQUID voltage
response marked by numbers I and II in Fig.~\ref{Fig1}b.

The first region (I in Fig.~\ref{Fig1}b) is in the vicinity of zero
external magnetic flux ($\phi_e \approx 0$) \cite{SQLin1}. In the
limit of zero SQUID inductance, $\beta_l = 0$, and for the bias
current equals the critical current, $i_b = 1$, SQUID voltage
response shape is described by the function: $u = |\sin\phi_e|$. For
the bias flux $\phi_b \leq \pi/2$ and inside the region $\phi_s \in
[-\phi_b/2, \phi_b/2]$ the voltage responses of DSQUID arms can be
written as $u_\pm = \pm\sin(\phi_s \pm\phi_b/2)$. Thus, in the range
where sine can be approximated by a linear function the total
response becomes linear:
\begin{equation}
u_ \Sigma \approx 2\phi_s\cos\phi_b/2.
\end{equation}

It is seen that the most linear part of the voltage response (at
$\phi_e = 0$) is moved from the boundary of SQUID working region ($0
\leq \phi_e \leq \pi/2$) to its center in DSQUID, making its
utilization possible. At the same time the bias flux providing
maximum transfer coefficient, $\phi_b = 0$, simultaneously makes the
width of the working (and linearized) region to be vanishing.

The second region (II in Fig.~\ref{Fig1}b) suitable for
linearization in DSQUID is located near the opposite boundary of
SQUID working region ($\phi_e \approx \pi/2$). Analytical
approximation of SQUID voltage response found in \cite{SQLin1,SQAn}
is
\begin{equation} \label{vresp}
u = u_0 - a\left[1 + (\beta_l^*u_0)^{-2} \right]^{-1}(i_b -
u_0)\tan^2\phi_e,
\end{equation}
where $u_0 = \sqrt{i_b^2 - \cos^2\phi_e}$ is the voltage response in
the limit of vanishing inductance, $\beta_l = 0$, and $a, \beta_l^*$
are parameters depended on $i_b, \beta_l$ (see Appendix).

\begin{figure}[]
\begin{center}
\includegraphics[bb = 0 0 296 155,width=1\columnwidth,keepaspectratio]{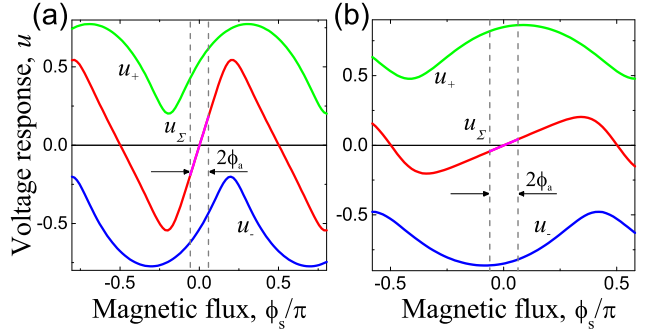}
\end{center}
\caption{DSQUID voltage response, $u_\Sigma$, and voltage responses
of its parts, $u_\pm$, for the following sets of parameters: (a)
$i_b = 1.02$, $\beta_l = 0.35\pi$, $\phi_b = 0.3895\pi$ and (b) $i_b
= 1.108$, $\beta_l = 0.35\pi$, $\phi_b = 0.8337\pi$. Boundaries of
the linear range of the total voltage response $u_\Sigma$ are marked
by vertical dotted lines. Linearity of the linear range is (a)
SFDR~=~92~dB, THD~=~$6\cdot 10^{-3}$\% and (b) SFDR~=~112~dB,
THD~=~$6 \cdot 10^{-4}$\%. The width of the linear range in both
cases corresponds to $2\phi_a \approx \pi/8$.} \label{Fig2}
\end{figure}

\begin{figure*}[]
\begin{center}
\includegraphics[bb = 0 0 519 282, width=1\textwidth,keepaspectratio]{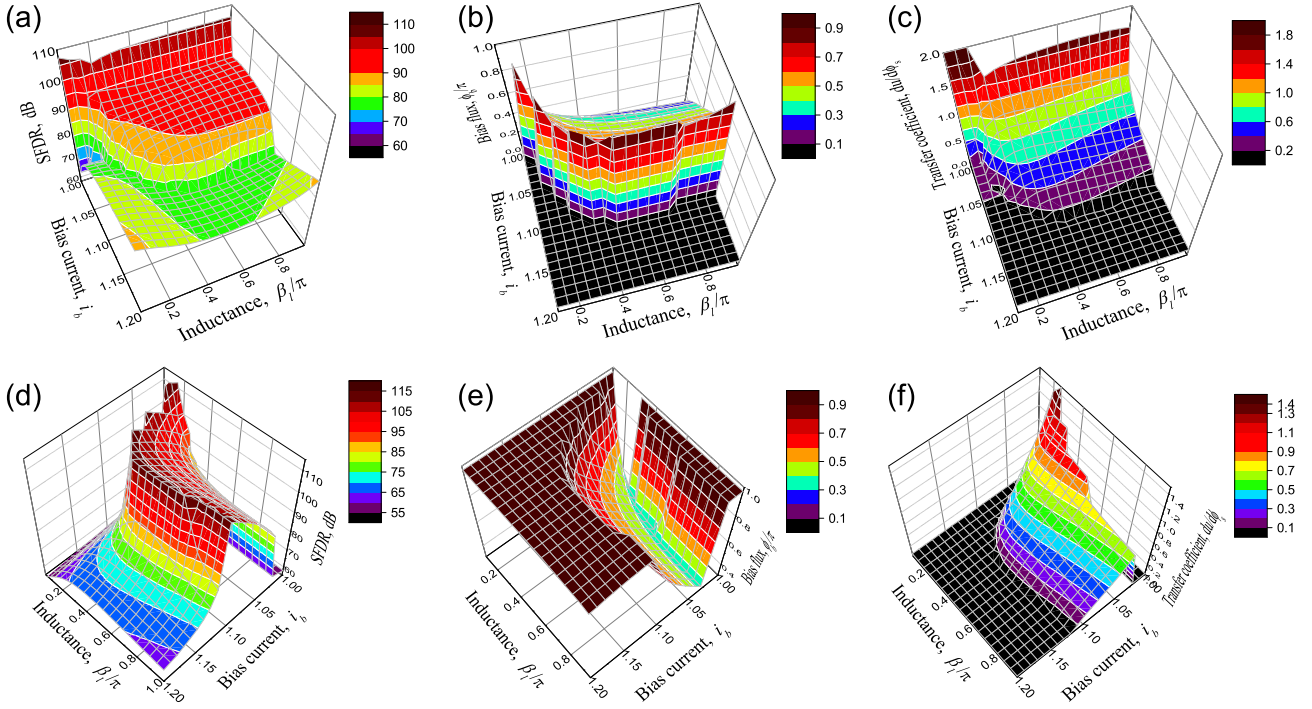}
\end{center}
\caption{Results of the bias flux optimization for obtaining the
highest SFDR of DSQUID voltage response under different conditions:
(i) $2\phi_a = 0.3\phi_b$ for (a), (b), (c) and (ii) $2\phi_a =
\pi/8$ for (d), (e), (f). SFDR is presented in (a), (d) planes, the
optimal bias flux $\phi_b/\pi$ is in (b), (e) planes, and the
transfer coefficient $du/d\phi_s$ is in (c), (f).} \label{Fig3}
\end{figure*}

In the vicinity of $\phi_e = \pi/2$ expression (\ref{vresp}) can be
accurately represented by a Taylor series limited to quadratic term:
\begin{equation} \label{Parabola}
\lim_{\phi_e \rightarrow \pi/2} u \approx i_b -
\frac{1}{2i_b}\left(Q + \left(1 - QZ\right)\left[\phi_e -
\frac{\pi}{2}\right]^2\right).
\end{equation}
Dependencies of $Q$ and $Z$ on $i_b, \beta_l$ are presented in the
Appendix.

For the bias flux close to $\phi_b = \pi$ voltage response of DSQUID
arms can be written as $u_\pm = u(\phi_s + \pi/2 \mp \delta)$, where
$\delta = \pi/2 - \phi_b/2$, due to symmetry of SQUID voltage
response. According to (\ref{Parabola}) this leads to linearized
total response:
\begin{equation}
u_\Sigma = \phi_s\frac{\pi - \phi_b}{i_b}(1 - QZ).
\end{equation}

While the width of the region II is defined by the range of validity
of approximation (\ref{Parabola}), the width of the linearized
region of DSQUID voltage response is determined by overlapping of
the regions II i.e. by the bias flux $\phi_b$. Deviation of the bias
flux from $\phi_b = \pi$ increases the transfer coefficient but
decreases the linearized region width.

Therefore, in both considered cases we face a tradeoff between the
transfer coefficient and the width of linearized region of DSQUID
voltage response. Below we show that it is possible to satisfy this
tradeoff for the bias flux values in the vicinity of $\phi_b = 0$ or
$\pi$. Corresponding examples shown in Fig.~\ref{Fig2} are discussed
in more detail below.

\section{Optimization of parameters}

Optimization procedure is performed in the range of the bias
current, $i_b \in [1,1.2]$, and the inductance, $\beta_l/\pi \in
[0.05, 1]$. Using standard Nelder-Mead simplex algorithm
\cite{MatlabNum} we numerically find the optimal bias flux providing
the highest SFDR of DSQUID voltage response.

\begin{figure}[]
\begin{center}
\includegraphics[bb = 0 0 768 588, width=1\columnwidth,keepaspectratio]{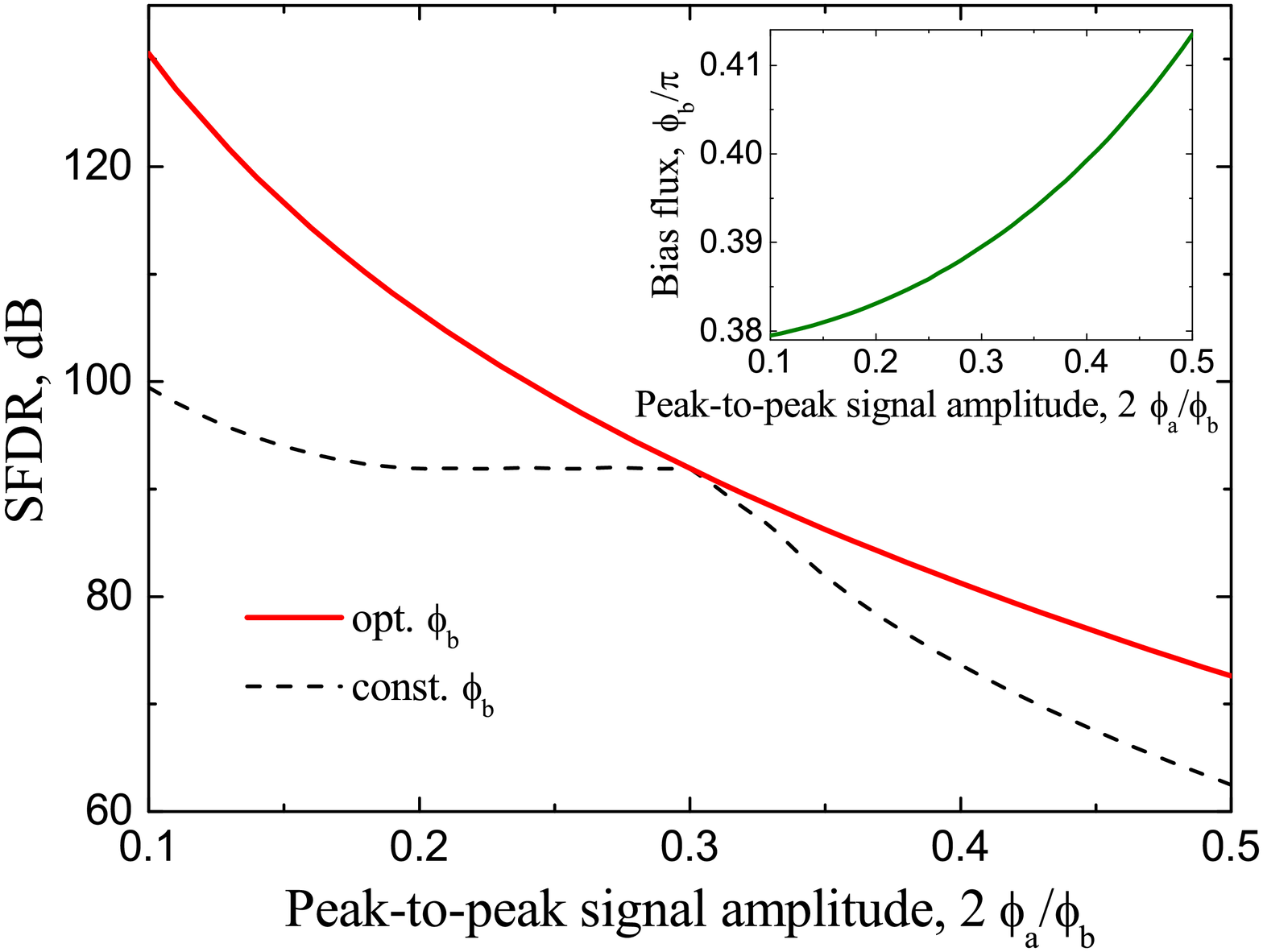}
\end{center}
\caption{SFDR of DSQUID voltage response versus peak-to-peak signal
amplitude for chosen values of parameters, $i_b = 1.02$, $\beta_l =
0.35\pi$, and optimal bias flux presented in inset (solid line).
SFDR for the constant bias flux, $\phi_b = 0.3895\pi$, is shown by
dotted line.} \label{Fig4}
\end{figure}

\begin{figure*}[]
\begin{center}
\includegraphics[bb = 0 0 567 141,width=1\textwidth,keepaspectratio]{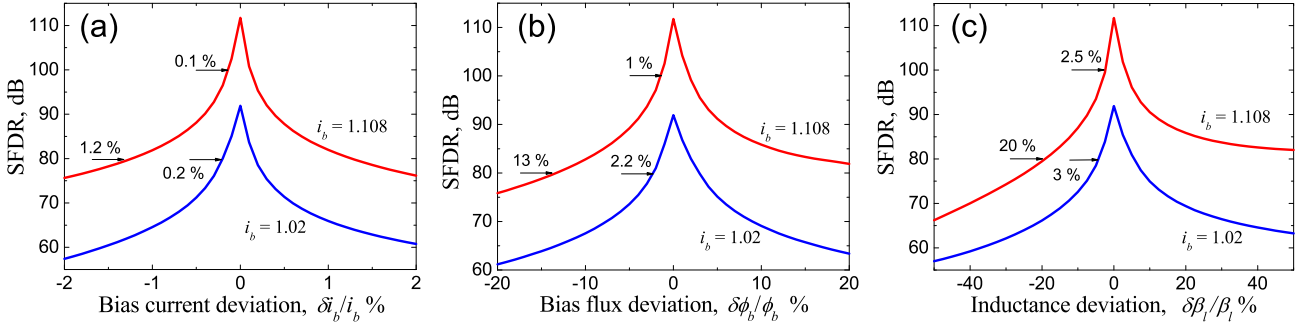}
\end{center}
\caption{SFDR of DSQUID voltage response versus deviation of (a) the
bias current, $\delta i_b/i_b$, (b) the bias flux, $\delta
\phi_b/\phi_b$, and (c) the inductance, $\delta \beta_l/\beta_l$,
for two chosen sets of parameters: (i) $i_b = 1.02$, $\beta_l =
0.35\pi$, $\phi_b = 0.3895\pi$ and (ii) $i_b = 1.108$, $\beta_l =
0.35\pi$, $\phi_b = 0.8337\pi$ (corresponding lines are marked by
$i_b$ values).} \label{Fig5}
\end{figure*}

We use two different conditions to consider linearization of the two
SQUID voltage response regions (I and II in Fig.~\ref{Fig1}b). The
first one is limitation of peak-to-peak signal to 30\% of the width
of DSQUID working region, $2\phi_a = 0.3\phi_b$. It is used to
consider utilization of the region I ($\phi_e \approx 0$). The usage
of the region II ($\phi_e \approx \pi/2$) is considered with
limitation of peak-to-peak signal to a fixed value equal to a
quarter of SQUID working region, $2\phi_a = \pi/8$.

Results of optimization obtained with utilization of the first
condition is presented in Fig.~\ref{Fig3}a,b,c. It is seen that the
highest SFDR~$>$~100~dB is obtained with the bias current close to
the critical current, $i_b \rightarrow 1$, see Fig.~\ref{Fig3}a.
However, while the transfer coefficient for this bias current is
also high (Fig.~\ref{Fig3}c), the bias flux is quite small
(Fig.~\ref{Fig3}b), and thus the width of DSQUID working region is
negligible.

Fortunately, there is a kind of plateau on $i_b, \beta_l$ plane of
parameters where $i_b > 1$ and linearity is still rather high,
SFDR~$>$~90~dB, see Fig.~\ref{Fig3}a. The example shown in
Fig.~\ref{Fig2}a corresponds to the values of parameters: $i_b =
1.02$, $\beta_l = 0.35\pi$ at the boundary of this plateau providing
both high enough bias flux, $\phi_b = 0.3895\pi$, and transfer
coefficient, $du/d\phi_s = 1.04$, while linearity is SFDR~=~92~dB
and THD~=~$6\cdot 10^{-3}$\%.

The chosen optimal bias flux corresponds to the width of the linear
range equal to approximately a quarter of the width of SQUID working
region, $2\phi_a \approx \pi/8$. The linearity increase is possible
with decrease of the signal amplitude as shown in Fig.~\ref{Fig4} by
solid line. However, the bias flux should be additionally slightly
tuned (see inset in Fig.~\ref{Fig4}). SFDR reaches 130~dB with
peak-to-peak signal equal to 10\% of the bias flux. SFDR obtained
with constant bias flux is presented by dotted line.

While limitation of signal amplitude to a percent of DSQUID working
region width makes utilization of the region I preferred, setting a
fixed signal amplitude vice versa allows the usage of the region II.
Optimization results obtained under utilization of the second
condition ($2\phi_a = \pi/8$) is presented in Fig.~\ref{Fig3}d,e,f.
Fig.~\ref{Fig3}d shows that for inductance values higher than
$\beta_l = 0.15\pi$ there are some values of the bias current $i_b >
1$ providing high linearity of the voltage response, SFDR~$>$~100~dB
(Fig.~\ref{Fig3}d), with the bias flux $\phi_b$ close to $\pi$
(Fig.~\ref{Fig3}e). Unfortunately, the transfer coefficient
decreases with increase of SFDR, see Fig.~\ref{Fig3}d,f. For the
previously chosen inductance, $\beta_l = 0.35\pi$, the optimal bias
current is $i_b = 1.108$ and with the bias flux equal to $\phi_b =
0.8337\pi$ the transfer coefficient is $du/d\phi_s = 0.22$.
Corresponding linearity is SFDR~=~112~dB, THD~=~$6 \cdot 10^{-4}$\%.
DSQUID voltage response for this set of parameters is presented in
Fig.~\ref{Fig2}b.

Fig.5 shows SFDR dependence on deviation of DSQUID parameters from
their optimal values, $\delta i_b/i_b$, $\delta \phi_b/\phi_b$,
$\delta \beta_l/\beta_l$, for both examples shown in
Fig.~\ref{Fig2}. It is seen that the linearity strongly depends on
the bias current, see Fig.5a. One should set $i_b$ with precision of
$\pm 0.1 - 0.2 \%$ to keep SFDR not less than 10~dB from its
maximum. At the same time for the same margin for SFDR the bias flux
$\phi_b$ can be set with an order less precision, $\pm 1 - 2 \%$,
see Fig.5b. Requirement for inductance value is even weaker
(Fig.5c).

\section{Discussion}

The data shown in Fig.~\ref{Fig5} indicate high requirements for the
identity of SQUIDs parameters in DSQUID. The statistics of Josephson
junctions critical currents fabricated by modern LTS fabrication
process can be described approximately as Gaussian with standard
deviation strongly depended on the junction size \cite{TolpStat}.
For the junctions with critical currents greater than 180~$\mu$A
(corresponding size is greater than 1500~nm) the standard deviation
is less than 1\% for the process using 248-nm photolithography, see
eq.(5) in \cite{Tolp}. Based on the data presented in
Fig.~\ref{Fig5}a, we can estimate the real attainable linearity
(SFDR) above 80 dB, accordingly.

Using conventional SQUID configuration one need restricting the
width of the working range below 0.1~$\Phi_0$ to make THD less than
1\% \cite{SQAmp}. With a similar width of the working region,
$2\Phi_a = \Phi_0/8$, DSQUID provides THD up to three orders better.
By assuming the root mean square flux noise of $\Phi_N =
10^{-6}$~$\Phi_0/$Hz$^{-1/2}$, one can obtain DSQUID dynamic range
of about 100~dB/Hz$^{-1/2}$. This means that with SFDR of an order
of 100~dB DSQUID allows truly linear magnetic flux-to-voltage
transformation since the distortions can be made lower than the
noise floor.

SQA can be based on a DSQUID as far as the DSQUID is the DQC. The
SQA formation can serve for the increase in dynamic range and
transfer coefficient \cite{SQA1}. HTS-based SQUID arrays recently
attracted attention due to the progress in fabrication of step-edge
junctions \cite{HTSSQIF} and the ones made with focused ion beam
\cite{HTSAnt,HTSFIB}. However, consideration of DSQUID-based array
requires complication of the model with taking into account the
noise and inductive coupling between cells \cite{SQLin2,BSQArr}.
Special attention should be paid to SQA matching with a load which
was thoroughly studied in \cite{DQCLoad}. It was shown that DQC
linearity is highly affected by the load and so high SFDR can be
obtained with a proper serial-parallel connection of the cells
keeping SQA impedance at least an order of magnitude lower than the
impedance of the load.

\section{Conclusion}

We considered differential SQUID - a ``DSQUID'' possessing highly
linear voltage response. The DSQUID is the differential connection
of two identical SQUIDs. Two regions of SQUID voltage response
suitable for linearization in DSQUID are identified. Arising
tradeoff between the transfer coefficient and the width of the
linearized region is revealed. Optimal values of the circuit
parameters providing high linearity of the voltage response
(SFDR~$>$~100~dB and THD~$<$~$10^{-3}\%$) are found. The linearity
dependence on deviation of the circuit parameters from their optimal
values is studied. It is shown that the ultimate linearity comes at
the cost of high identity of Josephson junctions (at the level of
tenths of a percent) and high-precision current supply (up to the
3-d decimal place).

\section{Acknowledgements}

This work was supported by Grant No. 17-12-01079 of the Russian
Science Foundation. Section 1 was written with the support of the
RFBR grant 16-29-09515 ofi$\_$m. V.I.R. acknowledges the Basis
Foundation scholarship.

\appendix
\section*{Appendix}

Expression (\ref{vresp}) is approximation of DC SQUID voltage
response shape. Analytical dependencies of its parameters $a$,
$\beta^*_l$ on $i_b$, $\beta_l$ are as follows \cite{SQLin1,SQAn}:
\begin{equation}
a = 2\kappa\nu(2i_b^2-1)/\xi,~~~\beta_l^*=\sqrt{\xi/\chi},
\end{equation}
where
\begin{subequations}
\begin{equation}
\kappa = \beta_l^{1.66}/(2.154\beta_l^{1.48} + 2.285),
\end{equation}
\begin{equation}
\nu = \beta_l^{1.92}/(4.28\beta_l^{1.625}+5.06),
\end{equation}
\begin{equation}
\xi = -0.586 \kappa + 2\nu + 4\kappa\nu(i_b^2 - 1),
\end{equation}
\begin{equation}
\chi = 0.586 i_b^2 \kappa - \nu - 2\kappa\nu(i_b^2 - 1).
\end{equation}
\end{subequations}

Parameters $Q, Z$ of the Taylor series (\ref{Parabola}) are
combinations of $a$, $\beta_l^*$ and $i_b$:
\begin{subequations}
\begin{equation}
Q = a\left[1+\left(\beta_l^*i_b\right)^{-2}\right]^{-1},
\end{equation}
\begin{equation}
Z = 1 -
\left[\left(\beta_l^*i_b\right)^2-3\right]\left(4i_b^2\left[1+
\left(\beta_l^*i_b\right)^2\right]\right)^{-1}.
\end{equation}
\end{subequations}

\section*{References}
\bibliographystyle{unsrt}
\bibliography{DSQL}

\begin{thebibliography}{10}

\bibitem{Tolp}
S.~K. Tolpygo.
\newblock Superconductor digital electronics: Scalability and energy efficiency
  issues.
\newblock {\em Low Temperature Physics}, 42(5):361--379, 2016.

\bibitem{HYPNP}
Niobium process.
\newblock http://www.hypres.com/foundry/niobium-process/.
\newblock Accessed: 2018-11-30.

\bibitem{OurBeil}
I.~I. Soloviev, N.~V. Klenov, S.~V. Bakurskiy, M.~Y. Kupriyanov, A.~L. Gudkov,
  and A.~S. Sidorenko.
\newblock Beyond {Moore's} technologies: Operation principles of a
  superconductor alternative.
\newblock {\em Beilstein Journal of Nanotechnology}, 8(1):2689--2710, 2017.

\bibitem{TolpLarge}
V.~K. Semenov, Y.~A. Polyakov, and S.~K. Tolpygo.
\newblock {AC}-biased shift registers as fabrication process benchmark circuits
  and flux trapping diagnostic tool.
\newblock {\em IEEE Transactions on Applied Superconductivity}, 27(4):7856973,
  2017.

\bibitem{HTSSQIF}
S.~Keenan, K.~Hannam, E.~Mitchell, J.~Lazar, C.~Lewis, A.~Grancea, W.~Purches,
  and C.~Foley.
\newblock Large high-temperature superconducting {2D} {SQIF} arrays.
\newblock In {\em European Conference on Applied Superconductivity (EUCAS),
  Geneva, Switzerland, 17-21 September 2017}, page EP172411, 2017.

\bibitem{SQLin2}
V.~K. Kornev, N.~V. Kolotinskiy, A.~V. Sharafiev, I.~I. Soloviev, and O.~A.
  Mukhanov.
\newblock Broadband active electrically small superconductor antennas.
\newblock {\em Superconductor Science and Technology}, 30(10):103001, 2017.

\bibitem{HTSSQIFA}
E.~E. Mitchell, K.~E. Hannam, J.~Lazar, K.~E. Leslie, C.~J. Lewis, A.~Grancea,
  S.~T. Keenan, S.~K.~H. Lam, and C.~P. Foley.
\newblock {2D} {SQIF} arrays using 20 000 {YBCO} high {Rn} {Josephson}
  junctions.
\newblock {\em Superconductor Science and Technology}, 29(6):06LT01, 2016.

\bibitem{HTSAnt}
F.~Couëdo, E.~R. Pawlowski, J.~Kermorvant, J.~Trastoy, D.~Crété,
  Y.~Lemaître, B.~Marcilhac, C.~Ulysse, C.~Feuillet-Palma, N.~Bergeal, and
  J.~Lesueur.
\newblock High-{Tc} superconducting antenna for highly-sensitive microwave
  magnetometry.
\newblock {\em arXiv}, page 1901.08786, 2019.

\bibitem{ADRHYP}
O.~A. Mukhanov, D.~Kirichenko, I.~V. Vernik, T.~V. Filippov, A.~Kirichenko,
  R.~Webber, V.~Dotsenko, A.~Talalaevskii, J.~C. Tang, A.~Sahu, P.~Shevchenko,
  R.~Miller, S.~B. Kaplan, S.~Sarwana, and D.~Gupta.
\newblock Superconductor digital-{RF} receiver systems.
\newblock {\em IEICE Transactions on Electronics}, E91-C(3):306--317, 2008.

\bibitem{IPHTSQ1}
V.~Zakosarenko, M.~Schulz, A.~Krueger, E.~Heinz, S.~Anders, K.~Peiselt, T.~May,
  E.~Kreysa, G.~Siringo, W.~Esch, M.~Starkloff, and H.~G. Meyer.
\newblock Time-domain multiplexed {SQUID} readout of a bolometer camera for
  {APEX}.
\newblock {\em Superconductor Science and Technology}, 24(1):015011, 2011.

\bibitem{Brag}
A.~I. Braginski.
\newblock Superconductor electronics: Status and outlook.
\newblock {\em Journal of Superconductivity and Novel Magnetism}, 32(1):23--44,
  2019.

\bibitem{SQA1}
V.~K. Kornev, A.~V. Sharafiev, I.~I. Soloviev, N.~V. Kolotinskiy, V.~A.
  Scripka, and O.~A. Mukhanov.
\newblock Superconducting quantum arrays.
\newblock {\em IEEE Transactions on Applied Superconductivity}, 24(4):06800001,
  2014.

\bibitem{SQA2}
V.~Kornev, A.~Sharafiev, I.~Soloviev, N.~Kolotinskiy, and O.~Mukhanov.
\newblock Superconducting quantum arrays for broadband {RF} systems.
\newblock In {\em Journal of Physics: Conference Series}, volume 507, page
  042019, 2014.

\bibitem{SQA3}
V.~K. Kornev, I.~I. Soloviev, A.~V. Sharafiev, N.~V. Klenov, and O.~A.
  Mukhanov.
\newblock Active electrically small antenna based on superconducting quantum
  array.
\newblock {\em IEEE Transactions on Applied Superconductivity}, 23(3):1800405,
  2013.

\bibitem{BSQ1}
V.~K. Kornev, I.~I. Soloviev, N.~V. Klenov, and O.~A. Mukhanov.
\newblock {Bi-SQUID}: A novel linearization method for dc {SQUID} voltage
  response.
\newblock {\em Superconductor Science and Technology}, 22(11):114011, 2009.

\bibitem{BSQ2}
V.~K. Kornev, I.~I. Soloviev, N.~V. Klenov, A.~V. Sharafiev, and O.~A.
  Mukhanov.
\newblock Linear {Bi-SQUID} arrays for electrically small antennas.
\newblock {\em IEEE Transactions on Applied Superconductivity}, 21(3 PART
  1):713--716, 2011.

\bibitem{BSQ3}
A.~Sharafiev, I.~Soloviev, V.~Kornev, M.~Schmelz, R.~Stolz, V.~Zakosarenko,
  S.~Anders, and H.~G. Meyer.
\newblock {Bi-SQUIDs} with submicron cross-type {Josephson} tunnel junctions.
\newblock {\em Superconductor Science and Technology}, 25(4):045001, 2012.

\bibitem{BSQ4}
V.~K. Kornev, A.~V. Sharafiev, I.~I. Soloviev, and O.~A. Mukhanov.
\newblock Signal and noise characteristics of {bi-SQUID}.
\newblock {\em Superconductor Science and Technology}, 27(11):115009, 2014.

\bibitem{BSQ5}
V.~K. Kornev, I.~I. Soloviev, N.~V. Klenov, and N.~V. Kolotinskiy.
\newblock Design issues of {HTS} {Bi-SQUID}.
\newblock {\em IEEE Transactions on Applied Superconductivity}, 26(5):7438807,
  2016.

\bibitem{BSQ6}
V.~K. Kornev, N.~V. Kolotinskiy, A.~Y. Levochkina, and O.~A. Mukhanov.
\newblock Critical current spread and thermal noise in {Bi-SQUID} cells and
  arrays.
\newblock {\em IEEE Transactions on Applied Superconductivity}, 27(4):7756342,
  2017.

\bibitem{BSQ7}
V.~K. Kornev, N.~V. Kolotinskiy, D.~E. Bazulin, and O.~A. Mukhanov.
\newblock High-inductance {Bi-SQUID}.
\newblock {\em IEEE Transactions on Applied Superconductivity}, 27(4):7752858,
  2017.

\bibitem{BSQ8}
V.~K. Kornev, N.~V. Kolotinskiy, D.~E. Bazulin, and O.~A. Mukhanov.
\newblock High-linearity {Bi-SQUID}: Design map.
\newblock {\em IEEE Transactions on Applied Superconductivity}, 28(7):1601905,
  2018.

\bibitem{DQC1}
V.~Kornev, N.~Kolotinskiy, V.~Skripka, A.~Sharafiev, I.~Soloviev, and
  O.~Mukhanov.
\newblock High linearity voltage response parallel-array cell.
\newblock In {\em Journal of Physics: Conference Series}, volume 507, page
  042018, 2014.

\bibitem{DQC2}
V.~K. Kornev, I.~I. Soloviev, N.~V. Klenov, and O.~A. Mukhanov.
\newblock Design and experimental evaluation of {SQIF} arrays with linear
  voltage response.
\newblock {\em IEEE Transactions on Applied Superconductivity}, 21(3 PART
  1):394--398, 2011.

\bibitem{DQC3}
V.~Kornev, I.~Soloviev, N.~Klenov, and O.~Mukhanov.
\newblock Progress in high-linearity multi-element {Josephson} structures.
\newblock {\em Physica C: Superconductivity and its Applications},
  470(19):886--889, 2010.

\bibitem{DQC4}
V.~K. Kornev, I.~I. Soloviev, N.~V. Klenov, and O.~A. Mukhanov.
\newblock High linearity {SQIF-like} {Josephson} junction structures.
\newblock {\em IEEE Transactions on Applied Superconductivity}, 19(3):741--744,
  2009.

\bibitem{DQC5}
V.~K. Kornev, I.~I. Soloviev, N.~V. Klenov, T.~V. Filippov, H.~Engseth, and
  O.~A. Mukhanov.
\newblock Performance advantages and design issues of {SQIFs} for microwave
  applications.
\newblock {\em IEEE Transactions on Applied Superconductivity}, 19(3):916--919,
  2009.

\bibitem{BSQex1}
G.~V. Prokopenko, O.~A. Mukhanov, A.~Leese De~Escobar, B.~Taylor, M.~C.
  De~Andrade, S.~Berggren, P.~Longhini, A.~Palacios, M.~Nisenoff, and R.~L.
  Fagaly.
\newblock {DC} and {RF} measurements of serial {Bi-SQUID} arrays.
\newblock {\em IEEE Transactions on Applied Superconductivity}, 23(3):6392883,
  2013.

\bibitem{BSQex2}
S.~Berggren, G.~Prokopenko, P.~Longhini, A.~Palacios, O.~A. Mukhanov, A.~L.
  de~Escobar, B.~J. Taylor, M.~C. de~Andrade, M.~Nisenoff, R.~L. Fagaly,
  T.~Wong, E.~Cho, E.~Wong, and V.~In.
\newblock Development of {2-D} {Bi-SQUID} arrays with high linearity.
\newblock {\em IEEE Transactions on Applied Superconductivity}, 23(3):6407797,
  2013.

\bibitem{diffSQUID}
D.~Drung, J.-H. Storm, and J.~Beyer.
\newblock {SQUID} current sensor with differential output.
\newblock {\em IEEE Transactions on Applied Superconductivity}, 23(3):1100204,
  2013.

\bibitem{SQLin1}
I.~I. Soloviev, N.~V. Klenov, A.~E. Schegolev, S.~V. Bakurskiy, M.~Y.
  Kupriyanov, M.~V. Tereshonok, and A.~A. Golubov.
\newblock Analytical description of low-{Tc} {DC} {SQUID} response and methods
  for its linearization.
\newblock In {\em 2017 16th International Superconductive Electronics
  Conference, ISEC 2017}, volume 2018-January, pages 1--3, 2018.

\bibitem{SQAn}
I.~I. Soloviev, N.~V. Klenov, A.~E. Schegolev, S.~V. Bakurskiy, and M.~Y.
  Kupriyanov.
\newblock Analytical derivation of {DC} {SQUID} response.
\newblock {\em Superconductor Science and Technology}, 29(9):094005, 2016.

\bibitem{MatlabNum}
J.~C. Lagarias, J.~A. Reeds, M.~H. Wright, and P.~E. Wright.
\newblock Convergence properties of the {Nelder-Mead} simplex method in low
  dimensions.
\newblock {\em SIAM Journal on Optimization}, 9(1):112--147, 1998.

\bibitem{TolpStat}
S.~K. Tolpygo, V.~Bolkhovsky, T.~J. Weir, L.~M. Johnson, M.~A. Gouker, and
  W.~D. Oliver.
\newblock Fabrication process and properties of fully-planarized deep-submicron
  {Nb/Al-AlOx/Nb} {Josephson} junctions for {VLSI} circuits.
\newblock {\em IEEE Transactions on Applied Superconductivity}, 25(3):1101312,
  2015.

\bibitem{SQAmp}
M.~Mueck and J.~Clarke.
\newblock Harmonic distortion and intermodulation products in the microstrip
  amplifier based on a superconducting quantum interference device.
\newblock {\em Applied Physics Letters}, 78(23):3666--3668, 2001.

\bibitem{HTSFIB}
B.~Mueller, M.~Karrer, F.~Limberger, M.~Becker, B.~Schroeppel, C.~J. Burkhardt,
  R.~Kleiner, E.~Goldobin, and D.~Koelle.
\newblock Josephson junctions and {SQUIDs} created by focused helium ion beam
  irradiation of {YBa2Cu3O7}.
\newblock {\em arXiv}, page 1901.08039v2, 2019.

\bibitem{BSQArr}
P.~Longhini, S.~Berggren, A.~L. de~Escobar, A.~Palacios, S.~Rice, B.~Taylor,
  V.~In, O.~A. Mukhanov, G.~Prokopenko, M.~Nisenoff, E.~Wong, and M.~C.~De
  Andrade.
\newblock Voltage response of non-uniform arrays of bi-superconductive quantum
  interference devices.
\newblock {\em Journal of Applied Physics}, 111:093920, 2012.

\bibitem{DQCLoad}
V.~K. Kornev, N.~V. Kolotinskiy, V.~A. Skripka, A.~V. Sharafiev, and O.~A.
  Mukhanov.
\newblock Output power and loading of superconducting quantum array.
\newblock {\em IEEE Transactions on Applied Superconductivity}, 25(3):1602005,
  2015.

\end{thebibliography}

\end{document}